\documentclass{emulateapj}

\shorttitle{Off-center HII Regions}
\shortauthors{Arthur}

\begin{document}
\title{Off-center HII regions  in power-law density distributions} 

\author{S. Jane Arthur}

\affil{Centro de Radioastronom\'\i{}a y Astrof\'\i{}sica,
  Universidad Nacional Aut\'onoma de M\'exico, Campus Morelia, Apartado Postal 3--72, 58090, Morelia,
  Michoac\'an, M\'exico }
\altaffiltext{1}{Email: j.arthur@astrosmo.unam.mx}

\begin{abstract}
The expansion of ionization fronts in uniform and spherically
  symmetric power-law density distributions is a well-studied
  topic. However, in many situations, such as a star formed at the
  edge of a molecular cloud core, an offset power-law density
  distribution would be more appropriate. In this paper a few of the
  main issues of the formation and expansion of HII regions in such
  media are outlined and results are presented for the particular
  cases where the underlying power laws are $r^{-2}$ and $r^{-3}$. A
  simple criterion is developed for determining whether the initial
  photoionized region will be unbounded, which depends on the
  power-law exponent and the ratio of the equivalent Str\"omgren
  radius produced by the star in a uniform medium to the stellar
  offset distance. In the expansion stage, the ionized volumes will
  eventually become unbounded unless pressure balance with the
  external medium is reached before the ionization front velocity
  becomes supersonic with respect to the ionized gas.
\end{abstract}


\section{Introduction}
\label{sec:intro}

The theoretical study of the formation and evolution of \ion{H}{2}
regions in non uniform media is motivated by the observational result
that the molecular clouds where massive star form generally possess
density gradients \citep[e.g.,][]{2007prpl.conf....3L}.  The assumed
density law is taken to be a power law $n \propto r^{-\alpha}$ with
exponents ranging from 1 to 3 \citep{1985ApJ...297..436A} obtained
from molecular line studies. Submillimeter continuum dust emission
observations of the molecular material around embedded ultracompact
\ion{H}{2} regions gives an exponent in the range 1.25 to 2.25
\citep{2003A&A...409..589H}. The most commonly found value for the
power-law exponent is $\alpha = 2$, which correponds to an isothermal,
self-gravitating sphere. However, steeper density laws, $\alpha < 4$
have been inferred from radio continuum spectra of a sample of
ultracompact \ion{H}{2} regions \citep{2000ApJ...544..277F}.

Numerical studies of \ion{H}{2} region expansion near the edge of a
molecular cloud (the so-called ``champagne flows''), were carried out
by Tenorio-Tagle and collaborators \citep{{1979A&A....71...59T},
{1979ApJ...233...85B}, {1981A&A....98...85B}}. In this model, an
\ion{H}{2} region formed near the edge of the cloud ``breaks out''
into the low density, intercloud medium during the expansion stage and
the ionized gas can achieve supersonic velocities of around
30~km~s$^{-1}$ in the accelerating flow. This problem was studied in
more detail by \citet{2005ApJ...627..813H}, and applied to the Orion Nebula
\ion{H}{2} region. They found that a long-lived quasi-stationary phase
exists in which the ionized flow is approximately steady in the frame
of reference of the ionization front. during this phase, the flow
structure is determined entirely by the distance of the ionizing star
from the front and the curvature of the front. The curvature
determines whether the flows are more ``champagne like'' or ``globule
like''. The curvature of the ionization front depends on the lateral
density distribution in the neutral gas.

\citet{{1989RMxAA..18...65F},{1990ApJ...349..126F}} developed
analytical solutions for the formation and expansion of \ion{H}{2}
regions in spherically symmetric, power-law density distributions,
where the ionizing star is assumed to be at the center of the density
distribution. The main result of their paper is that for spherical
clouds with power-law density distributions $r^{-\alpha}$, there is a
critical value of the exponent, $\alpha_{\rm crit} = 3/2$ above which
the cloud becomes completely ionized during the expansion phase of the
\ion{H}{2} region. For power-law exponents higher than this critical
value, two regimes were identified: a slow regime, in which the
expansion of the dense cloud core is mildly supersonic and has almost
constant velocity, corresponding to $3/2 < \alpha \le 3$, and the fast
regime, where the core expansion drives a strongly accelerating shock,
which corresponds to $\alpha > 3$. This problem was revisited by
\citet{2002ApJ...580..969S}, who developed a self-similar solution for
the internal motions in the ionized gas in the $\alpha > 3/2$ cases.

\citet{{1989RMxAA..18...65F},{1990ApJ...349..126F}} also considered
self-gravitating isothermal disk density distributions where the
density fall-off is steeper than $r^{-3/2}$. In this case, the
\ion{H}{2} region becomes unbounded in a conical section of the disk,
where the opening angle of the cone depends on the ratio of the
Str\"omgren radius in the midplane density to the density distribution
scale height. During the expansion stage of such \ion{H}{2} regions,
the flattening of the internal ionized gas density distribution can
lead to the trapping of an initially unbounded ionization front.

More recently, photoionized regions in a variety of axisymmetric
density distributions were modeled numerically by
\citet{2006ApJS..165..283A}, who also included the central star's
stellar wind. In that work, it was found that the hot bubble created
by the shocked stellar wind results in a champagne flow being set up
outside the swept-up stellar wind shell, leading to mixed kinematics
from an observational point of view. \citet{2006ApJS..165..283A} also
considered the possibility of stellar motion up the density gradient,
which introduces the added complication of bowshock kinematics to the
problem.

\citet{2006arXiv.0605501.M} performed low-resolution three-dimensional
numerical simulations of \ion{H}{2} region dynamical evolution in a
collapsing molecular cloud. Here, the density distribution had a
core-envelope structure, where the envelope had a power-law ($r^{-2}$)
density distribution. Both turbulent and non-turbulent cases were
modeled. In the non-turbulent case, even for ionizing sources
off-center from the core, the \ion{H}{2} region expansion was nearly
spherical for the parameters adopted and the timescale of the
simulation.

The photoionized regions produced by moving stars were studied
numerically by \citet{2005arXiv.0508467.F}. In that paper, the
\ion{H}{2} regions produced by a star off-center in a spherically
symmetric density distribution were modeled, where a range of stellar
velocities (between 0 and 12~km~s$^{-1}$) were considered. It was
found that the \ion{H}{2} region produced by a star off-center in an
underlying $r^{-2}$ density distribution did not become unbounded,
while that in a $r^{-3}$ density distribution did, for the case of a
stationary star. This appears to contradict the main result of
\citet{1990ApJ...349..126F}. This result is the motivation for the
present work, where we aim to provide simple tools for analyzing the
formation and expansion of \ion{H}{2} regions off-center in
spherically symmetric density distributions.

The structure of this paper is as follows: in \S~\ref{sec:prob} we
outline the problem and discuss the initial formation stage of an
\ion{H}{2} region off-center in a spherically symmetric density
distribution and find a criterion for the \ion{H}{2} region to remain
bounded. In \S~\ref{sec:expand} we investigate the expansion stage and
derive a simple differential equation that describes the expansion of
the ionization front along the symmetry axis. We assess the validity
of our assumptions about the internal structure of the ionized gas by
comparing our analytical model with the results of numerical
simulations in \S~\ref{sec:numerical}. Finally, in
\S~\ref{sec:summary} we summarize our results.

\section{Description of the problem}
\label{sec:prob}

\begin{figure}[!t]
\includegraphics[width=0.45\textwidth]{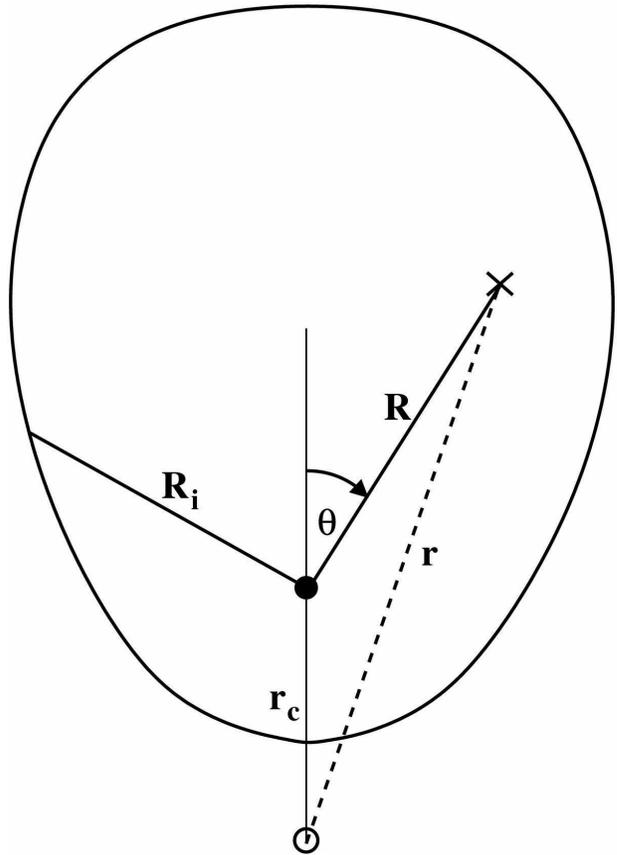}
\caption{Geometrical configuration. The center of the (spherically
  symmetric) power-law density
  distribution is denoted by the open circle, the star is represented
  by the filled circle, a distance $r_c$ from the density center. For
  a parcel of gas at X, distance $r$ from the density center, the
  star-centered coordinates are $(R,\theta)$, where $R$ is the
  distance from the star and $\theta$ is measured clockwise from the
  extension of the line joining the star to the density
  center. Rotational symmetry is assumed around the axis joining the
  star to the center of the density distribution. The
  position of the ionization front in this coordinate system is
  $R_i$. An approximate shape for the ionization front is also shown.}
\label{fig:geom}
\end{figure}

The underlying, spherically symmetric density distribution is
described by 
\begin{equation}
n(r) = n_c\left(\frac{r}{r_c}\right)^{-\alpha}
\label{eq:denorig}
\end{equation}
where $n_c$ is the number density at the position of the star,
$r_c$. We now change our reference system and rewrite the density
distribution in axisymmetric coordinates centered on the star
position, with the distance from the star, $R$, found from
\begin{equation}
  R^2 - 2r_cR\cos(\pi-\theta) + r_c^2 - r^2 = 0 \ ,
\end{equation}
where $\theta$ is measured from the symmetry axis and the center of
the density distribution is in the direction $\theta = \pi$, see
Figure~\ref{fig:geom}. The density distribution of
Equation~\ref{eq:denorig}, in terms of this new reference system,
$(R,\theta)$, becomes
\begin{equation}
  n(R,\theta) = n_c \left( \frac{R^2 + r_c^2 + 2r_cR\cos
  \theta}{r_c^2} \right)^{-\alpha/2} \ .
\label{eq:denrtheta}
\end{equation}
Along the symmetry axis, $\theta = 0$, this can be written
\begin{equation}
  n(y) = n_c \left( 1 + y \right)^{-\alpha} \ ,
\label{eq:denyplus}
\end{equation}
where we have defined $y = R/r_c$.

For $y < 1$, expansion to linear terms in a Taylor series shows that
\begin{equation}
  n(y) \approx n_c (1 - \alpha y) \ ,
\label{eq:densmally}
\end{equation}
which is no longer a power law. For $y > 1$ we find $n(y) \approx n_c
y^{-\alpha}$, that is, the offset density distribution resembles the
original power-law density distribution. Consequently, if the initial
Str\"omgren radius, $R_s$, is smaller than the stellar offset
distance, then we expect the initial \ion{H}{2} region to remain
confined, even for power laws steeper than $\alpha > 3/2$.

\subsection{Condition for the initial Str\"omgren region}
\label{sec:condition}
We can develop a quantitative criterion for deciding whether the
initial photoionized region will remain bounded in an offset power-law
density distribution by balancing recombinations and ionizations,
assuming pure hydrogen, along the symmetry axis $\theta = 0$:
\begin{equation}
  \alpha_B r_c^3 \int^y_0 n_c^2 \left( 1 + y^\prime \right)^{-2\alpha}
  y^{\prime 2} \ dy^\prime = \frac{Q_H}{4\pi} \ , 
\label{eq:ionbalance}
\end{equation}
where $\alpha_B$ is the case B recombination coefficient. This
equation can be written
\begin{equation}
  \int^y_0 \left( 1 + y^\prime \right)^{-2\alpha} y^{\prime 2} \ dy^\prime
  = \frac{1}{3}y^3_{\rm sc} \ ,
\label{eq:iontidy}
\end{equation}
where $y_{\rm sc}$ is the non-dimensionalized Str\"omgren radius
for a uniform medium of number density $n_c$, i.e.,
\begin{equation}
  y_{\rm sc} = \frac{R_s}{r_c} = \left( \frac{3 Q_H}{4\pi \alpha_B
  n_c^2 r_c} \right)^{1/3} \ .
\label{eq:nondimstrom}
\end{equation}
In the ``uphill'' direction, $\theta = \pi$, the equivalent
non-dimensionalized ionization balance equation is
\begin{equation}
  \int^y_0 \left(1 - y^\prime\right)^{-2\alpha} y^{\prime 2} \ dy^\prime =
  \frac{1}{3} y_{\rm sc}^3 \ .
\label{eq:uphill}
\end{equation}
Equations~\ref{eq:nondimstrom} and \ref{eq:uphill} define integral
equations for the extent of the ionized 
region, $y$, along $\theta = 0$ and $\theta = \pi$, respectively, as
functions of $y_{\rm sc}$ and $\alpha$.
\begin{figure}[!t]
\includegraphics[width=0.45\textwidth]{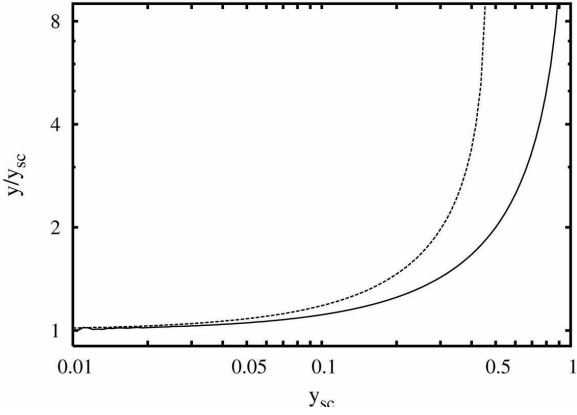}
\caption{Ratio of ionization front radius along $\theta = 0$ to
  uniform density Str\"omgren radius, $y/y_{\rm sc}$, as a function of
  uniform density Str\"omgren radius, $y_{\rm sc}$, for power-law density
  distributions with exponent $\alpha = 2$ (\textit{solid line}) and $\alpha =
  3$ (\textit{dashed line}). The asymptotes are $y_{\rm sc} = 1$ for $\alpha =
  2$ and $y_{\rm sc} = 0.464$ for $\alpha = 3$. }
\label{fig:newt}
\end{figure}

Integrating Equation~\ref{eq:iontidy} for $\alpha > 3/2$ we obtain
\begin{eqnarray}
  \frac{1}{3} y_{\rm sc}^3 = &
 \frac{-2}{(1-2\alpha)(2 -
  2\alpha)(3-2\alpha)}(1 - [1+y]^{3-2\alpha}) \nonumber \\
 & - \frac{2}{(1-2\alpha)(2-2\alpha)}(y[1+y]^{2-2\alpha}) \nonumber \\
 & + \frac{1}{(1-2\alpha)}(y^2[1+y]^{1-2\alpha}) \ .
\label{eq:intresult}
\end{eqnarray}
In general, this yields a positive, real root for $y$ (the
non-dimensionalized ionization front radius) as long as
\begin{equation}
\frac{1}{3}y_{\rm sc}^3 < \frac{2}{(2\alpha-1)(2\alpha-2)(2\alpha-3)}
\ .
\label{eq:root}
\end{equation}
For example, when $\alpha = 2$, the requirement is $y_{\rm sc} < 1$, while for
$\alpha = 3$, the condition is $y^3_{\rm sc} < 0.1$, i.e., $y_{\rm sc}
< 0.464$, in order that there be a real,
positive root for $y$. In Figure~\ref{fig:newt} we plot the ratio
$y/y_{\rm sc}$ against $y_{\rm sc}$ for $y_{\rm sc}$ values between
0.01 and 1.0 for the two examples above, $\alpha = 2$ and $\alpha = 3$.
We see that for $y_{\rm sc}$ less than the critical value, the value
of the root $y$ is not too different from $y_{\rm sc}$, but close to
the critical value $y/y_{\rm sc} \rightarrow \infty$.

\section{Expansion stage}
\label{sec:expand}
In order to study the expansion of the ionized gas after the initial
formation of the photoionized region we follow the same analysis as
\citet{1968dms..book.....S} and \citet{1992phas.book.....S}. As the
\ion{H}{2} region expands, a neutral shock is sent out ahead of the
ionization front into the ambient medium.

Across the isothermal shock, the relevant equations are mass conservation
\begin{equation}
  \rho_{1s} v_{1s} = \rho_{2s} v_{2s} \ ,
\label{eq:cont}
\end{equation}
and momentum conservation
\begin{equation}
  p_{1s} + \rho_{1s} v_{1s}^2 =   p_{2s} + \rho_{2s} v_{2s}^2
\label{eq:mom}
\end{equation}
where the subscript $1s$ refers to preshock
conditions and $2s$ refers to postshock conditions and all velocities
are in the frame of the shock. We also have the equation
for the density jump across a
weak-D ionization front
\begin{equation}
\frac{\rho_{2i}}{\rho_{1i}} = \frac{1}{2c_2^2} \left\{%
  \left(c_1^2 + v_{1i}^2 \right)^2 + \left[ \left( c_1^2 + v_{1i}^2
  \right)^2 - 4c_2^2 v_{1i}^2 \right]^{1/2} \right\}
\label{eq:ifront}
\end{equation}
where the subscript $1i$ refers to conditions conditions in front of
the ionization front and $2i$ refers to conditions in the photoionized
gas just behind the ionization front. Here, $c_1$ is the sound speed
in the neutral gas and $c_2$ is the sound speed in the ionized gas,
and $v_{1i}$ refers to the velocity of the gas ahead of the ionization
front in the frame of the ionization front.

Furthermore, we assume that the region between the shock front and the
ionization front is thin, i.e. $u_{1i} = u_{2s}$, where $u_{1i}$ and
$u_{2s}$ are the velocities of the gas ahead of the ionization front
and behind the shock front in the frame of the star, and that the
density of the ionized gas inside the photoionized region is spatially
uniform but decreases with time, i.e. $\rho_{2i} \equiv
\rho_{2i}(t)$. This is valid if the sound-crossing time of the ionized
region is short compared to the expansion timescale.

As a first approximation to the non-spherical shape of the ionization
front in a non-uniform density distribution, we assume that the radius
of the ionization front can be described by
\begin{equation}
  R_i = a_0 + a_1 \cos \theta \equiv a_0 + a_1 \mu
\label{eq:radius}
\end{equation}
to first order in $\mu = \cos\theta$, where 
\begin{equation}
  a_0 = \frac{R_- + R_+}{2}
\label{eq:a0}
\end{equation}
and
\begin{equation}
  a_1 = \frac{R_- - R_+}{2}
\label{eq:a1}
\end{equation}
with $R_-$ corresponding to the shell radius when $\mu = -1$ and $R_+$
to the case $\mu = 1$ \citep[see, e.g.,][]{1977A&A....59..161D}.

The volume contained within the ionization front $R_i$ is 
\begin{equation}
  V_i = \frac{2\pi}{3} \int_{-1}^{1} R(\mu)^3 \ d\mu \ ,
\label{eq:volume}
\end{equation}
which in this approximation becomes
\begin{equation}
  V_i = a_0^3 \left(1 + \frac{a_1^2}{a_0^2} \right) \ .
\label{eq:volume2}
\end{equation}

Assuming uniform expansion, then the mass contained within a
volume, $V$, interior to an internal  shell of
radius $R$ must be conserved during the expansion. That is,
\begin{equation}
  \rho_{2i} V = \mbox{constant} \ ,
  \label{eq:shellmass}
\end{equation}
and the total volume of ionized gas satisifies the ionization balance
equation, where the ionizing photon output of the star is assumed to
be constant, that is
\begin{equation}
  \rho_{2i}^2 V_i = \mbox{constant} \ .
\label{eq:ibalance}
\end{equation}
Taking the time derivatives of Equations~\ref{eq:shellmass} and
\ref{eq:ibalance} we find that
\begin{equation}
  \left. \frac{dV}{dt} \right|_{R_i} = \frac{1}{2} \frac{dV_i}{dt} \ .
\label{eq:volrelate}
\end{equation}
where we have evaluated the derivative for a shell located just inside
the ionization front radius.
Writing 
\begin{equation}
  \frac{dV}{dt} = \frac{dV}{dR}\frac{dR}{dt} \mbox{\hspace*{2em} and
  \hspace*{2em}} \frac{dV_i}{dt} = \frac{dV_i}{dR_i}\frac{dR_i}{dt} 
\label{eq:dvdt}
\end{equation}
then we find that the gas velocity behind the ionization front is half
the ionization front velocity, $u_{2i} = 0.5 U_i$, where $u_{2i} =
\left. \frac{dR}{dt} \right|_{R_i}$ and $U_i = \frac{dR_i}{dt}$.
This assumes that the expansion is uniform and that there are no lateral gas
motions within the photoionized region.

With the additional assumption that $v_{1i}/{c_1} \ll 1$, then
equations~\ref{eq:cont}, \ref{eq:mom}, \ref{eq:ifront}, together with
$u_{2i} = 0.5 U_i$, can be combined to give
\begin{equation}
  \frac{\rho_{2i}}{\rho_{1s}} = \frac{4U_i^2}{4c_2^2 + U_i^2} \ ,
\label{eq:denrat}
\end{equation}
where we have used the thin shell approximation $U_i \approx U_s$.
These assumptions are the same as those of \citet{1968dms..book.....S}.

The density ratio can be found from ionization balance
\begin{equation}
  \rho_{2i}^2 V_i = \rho_c^2 R_0^3 \ ,
\label{eq:ibalance1}
\end{equation}
where $R_0$ is the initial Str\"omgren radius formed in a medium of
uniform mass density $\rho_c$. We also have the offset power-law
initial density distribution, written here for $\mu \pm 1$:
\begin{equation}
  \rho_{1s} = \rho_c \left(1 \pm \frac{R}{r_c} \right)^{-\alpha} \ ,
\label{eq:offset}
\end{equation}
where $r_c$ is the offset distance of the star from the center of the
density distribution and $\rho_c$ is the density at this
point. Thus
\begin{equation}
  \frac{\rho_{2i}}{\rho_{1s}} = \left( \frac{R_0^3}{V_i}\right)^{1/2}
  \left(1 \pm \frac{R}{r_c} \right)^{\alpha} \ 
\label{eq:denrat2}
\end{equation}
in the directions $\mu = \pm 1$ along the symmetry axis.

Equation~\ref{eq:denrat} can be written in the form
\begin{equation}
  M^2 \left(4 - \frac{\rho_{2i}}{\rho_{1s}} \right) = 4
  \frac{\rho_{2i}}{\rho_{1s}} \ ,
\label{eq:emsq}
\end{equation}
where $M = U_i/c_2$. This constitutes a differential equation for the
radius of the ionization front, $R_i$, as a function of time and can
be solved numerically with standard techniques
\citep[e.g.,][]{1992nrfa.book.....P}, by considering directions $\mu =
1$ and $\mu = -1$ together and using equations~\ref{eq:radius},
\ref{eq:a0}, \ref{eq:a1}, \ref{eq:volume} and \ref{eq:denrat2}, with
initial conditions $M = 2/\sqrt{3}$ when $t = 0$, $R_i = R_0$ and
$\rho_{2i} = \rho_{1s}$. Here, it is assumed that $R_0$ is much less
than the critical value found from solving Equation~\ref{eq:root}
above, and so the initial ionized region radius ($R_i$ at $t = 0$) is
taken to be that for a uniform medium.
\begin{figure}[!t]
\includegraphics[width=0.45\textwidth]{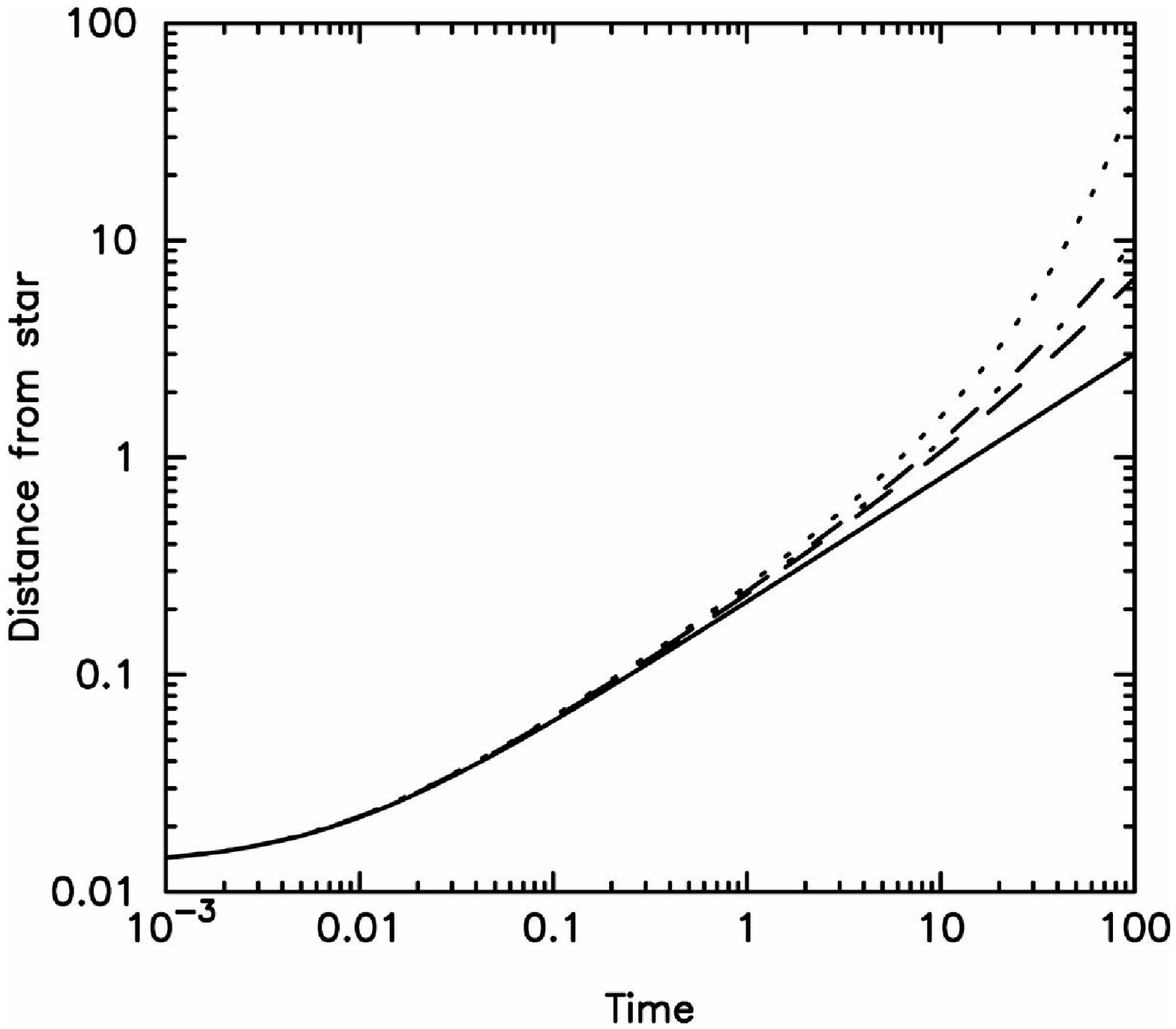}\\
\includegraphics[width=0.45\textwidth]{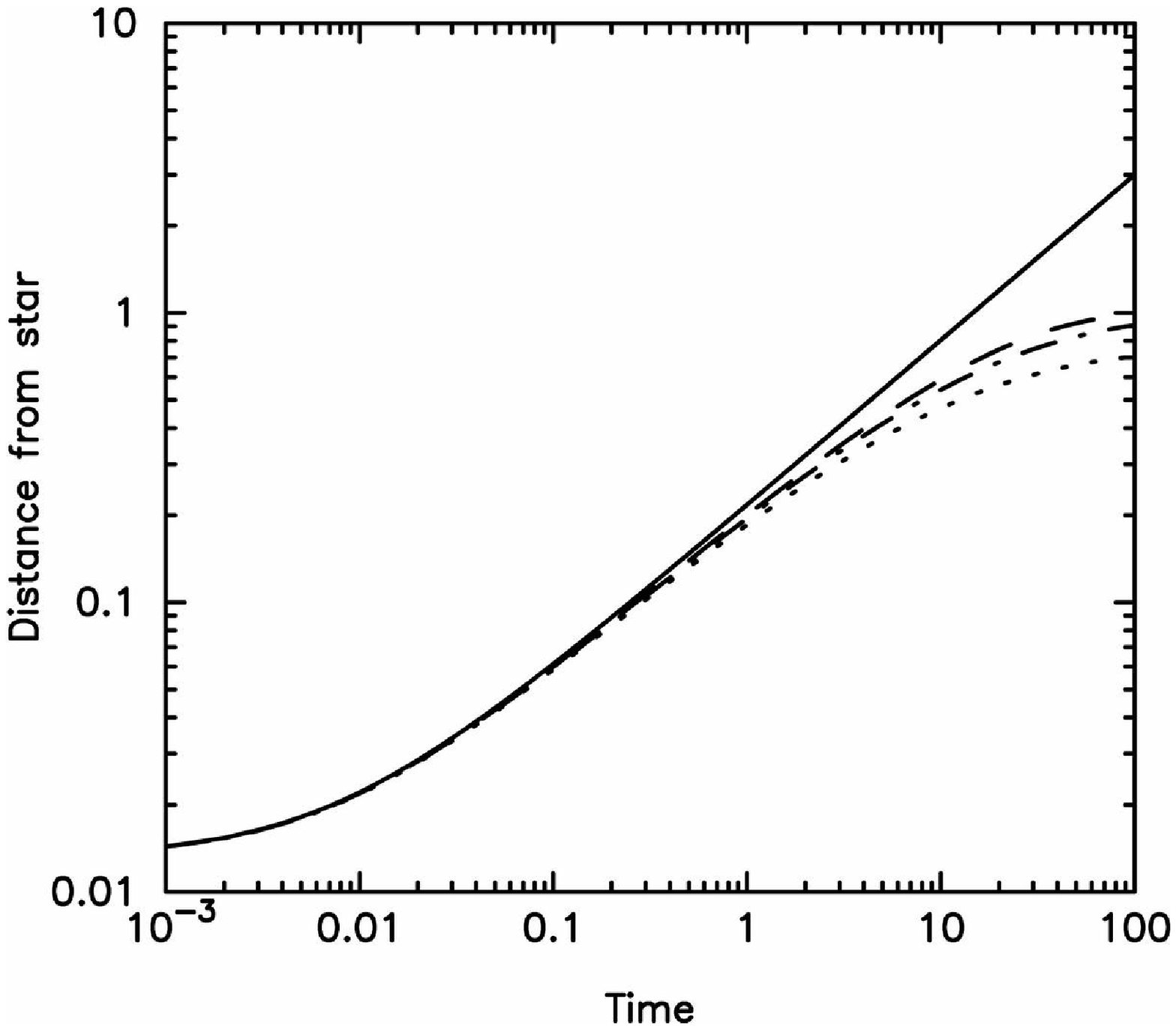}
\caption{Radius of the ionization front (in units of the stellar
  offset radius, $r_c$) against time (in units of the sound crossing
  time, $\tau = r_c/c_2$). \textit{Top}: $\mu = 1$ direction.
  \textit{Bottom}: $\mu = -1$ direction. Density power laws are
  $\alpha = 0$ (\textit{solid line}), $\alpha = 3/2$ (\textit{dashed
  line}), $\alpha = 2$ (\textit{dot-dashed line}), and $\alpha = 3$
  (\textit{dotted line}).}
\label{fig:radupdown}
\end{figure}
\begin{figure}[!t]
\includegraphics[width=0.45\textwidth]{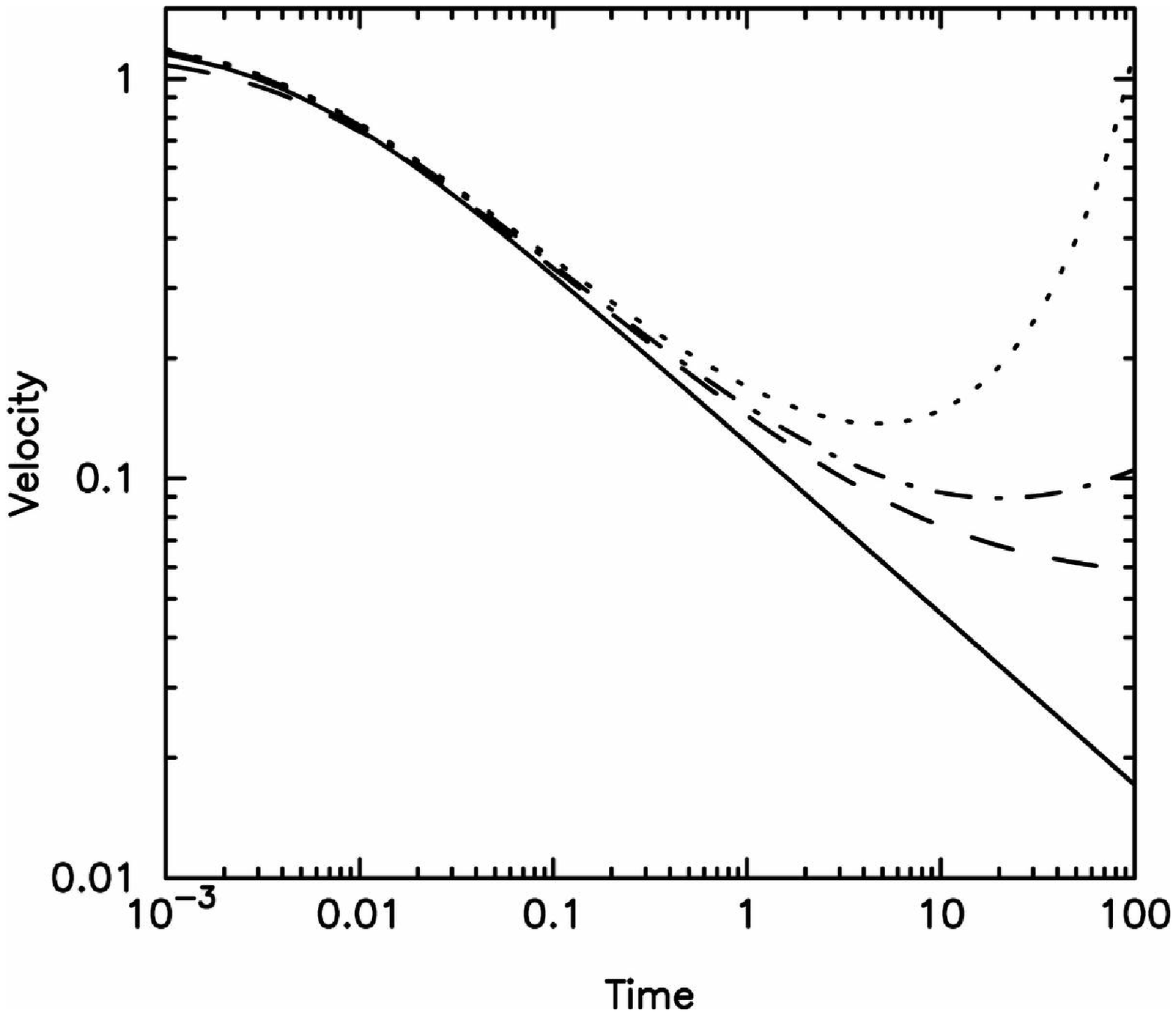}\\
\includegraphics[width=0.45\textwidth]{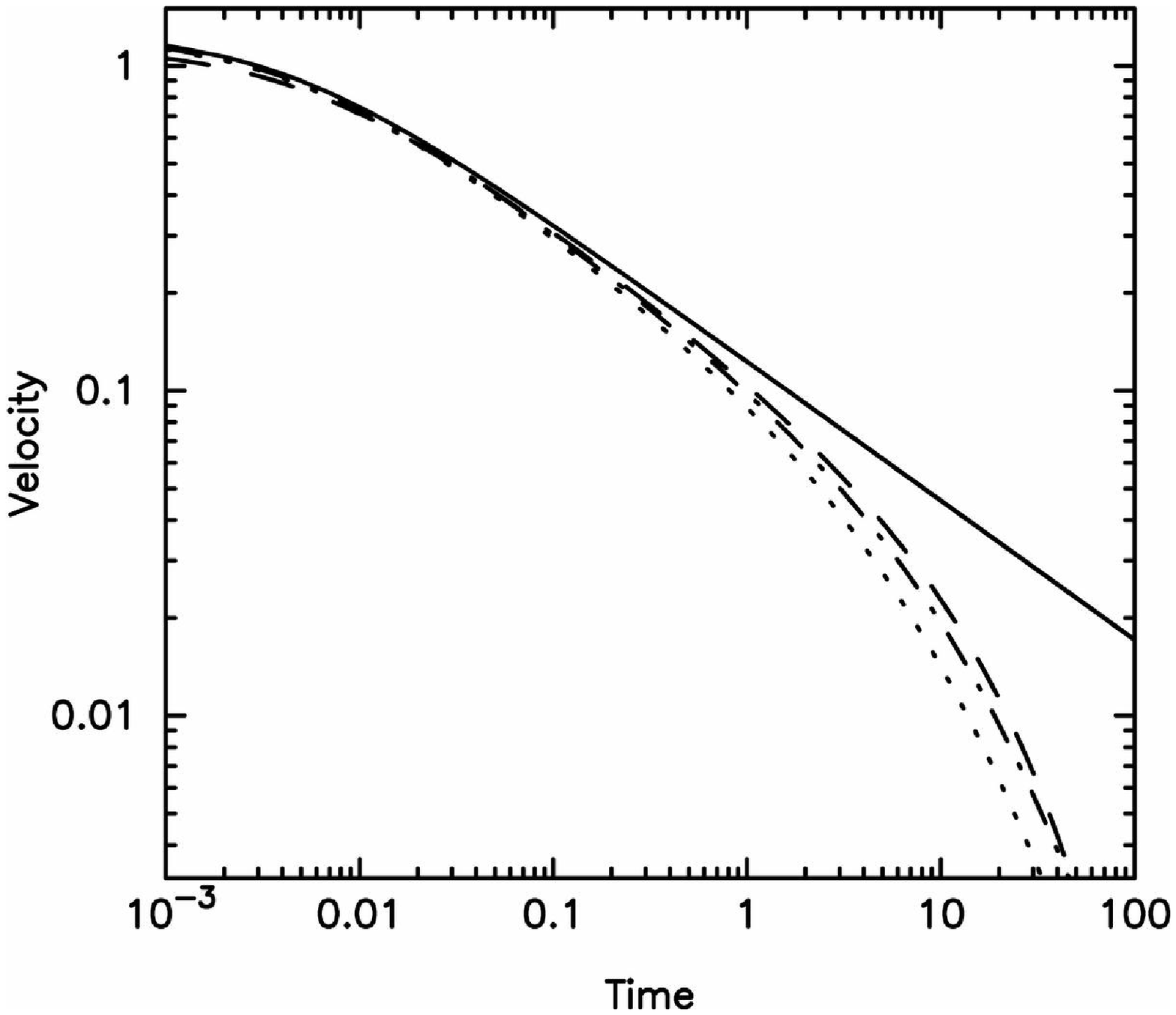}
\caption{Velocity of the ionization front (in units of the ionized gas
  sound speed) against time (in units of the sound crossing time).
  \textit{Top}: $\mu = 1$ direction. \textit{Bottom}: $\mu = -1$
  direction. Linetypes same as in Fig.~\ref{fig:radupdown}.}
\label{fig:velupdown}
\end{figure}

In Figures~\ref{fig:radupdown} and \ref{fig:velupdown} we can see the
result of solving Equation~\ref{eq:emsq} numerically for the cases
$\alpha = 0$, 2 and 3. In these figures, all distances are scaled to
$r_c$ and times are scaled with the sound crossing time of this
distance in the ionized gas, $\tau = r_c/c_2$. The results presented
are particular to the case $y_{\rm sc} = R_0/r_c = 0.01324$, which
corresponds to an ionizing source with $Q_H = 10^{48}$
photons~s$^{-1}$ and a density $n_c = 10^7/e$~cm$^{-3}$, where we have
also assumed that the case B recombination coefficient takes the value
$\alpha_B = 2.6 \times 10^{-13}$~cm$^3$~s$^{-1}$. These values of the
parameters are motivated by \citet{2005arXiv.0508467.F}.

In the ``upward'' direction ($\mu = -1$, density increasing), as
expected the radius grows more slowly in the $\alpha = 2$ and $3$
cases than for the uniform medium, classical case $\alpha = 0$, as
shown in Figures~\ref{fig:radupdown} and \ref{fig:velupdown}. In the
``downward'' direction, $\mu = 1$, Figures~\ref{fig:radupdown} and
\ref{fig:velupdown} clearly show that in the $\alpha = 3$ case, the
velocity starts to increase after about 5 sound crossing times and by
100 sound crossing times the velocity of the ionization front has
become supersonic with respect to the ionized gas. This means that the
ionization front has overtaken the shock wave and will ionize out to
infinity. In the $\alpha = 2$ case, although the velocity does start
to increase again after about 13 sound crossing times, the increase is
only very gradual, and within the timescale studied the ionization front
does not become supersonic with respect to the ionized gas.

\begin{figure}[!t]
\includegraphics[width=0.45\textwidth]{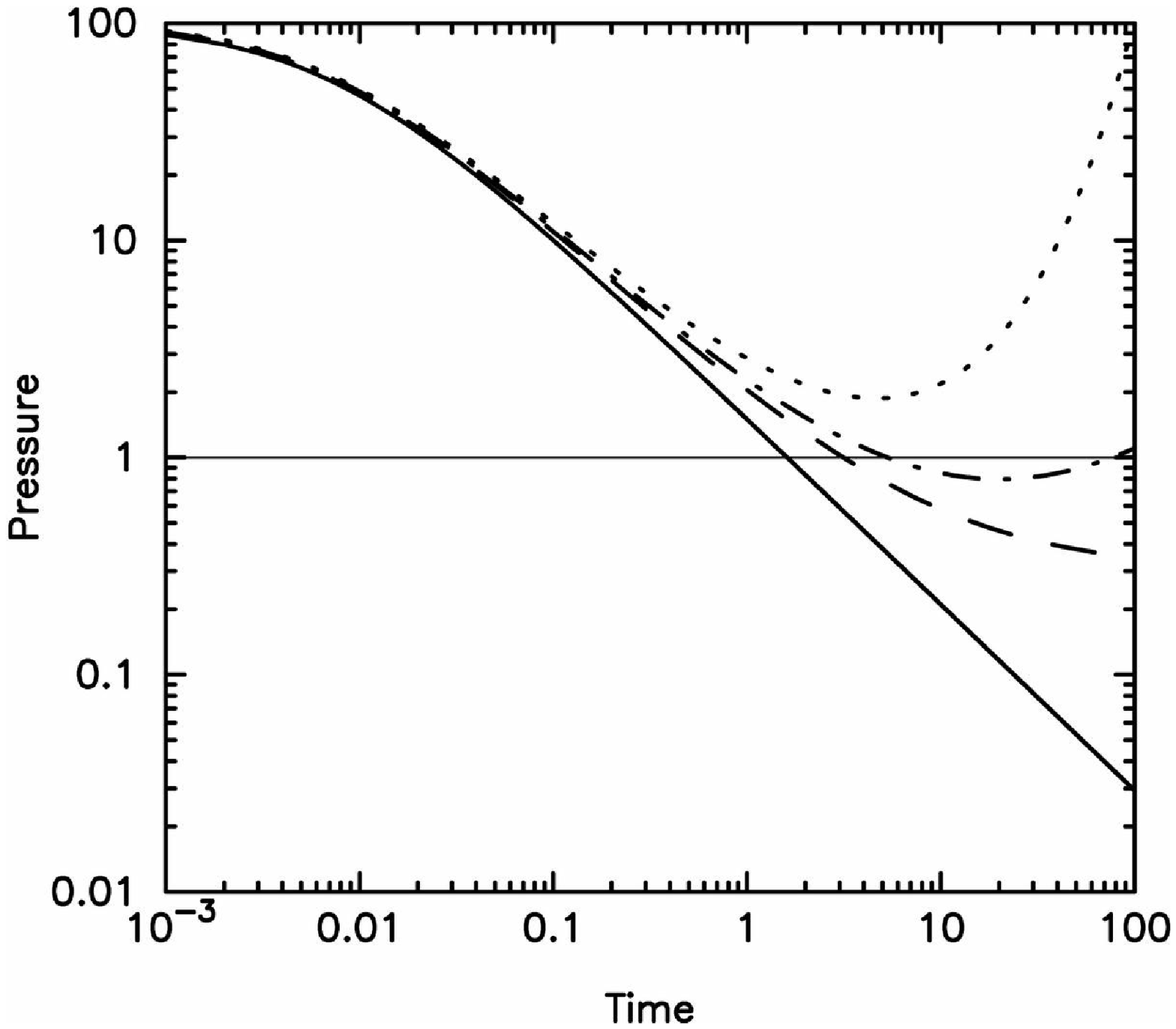}\\
\includegraphics[width=0.45\textwidth]{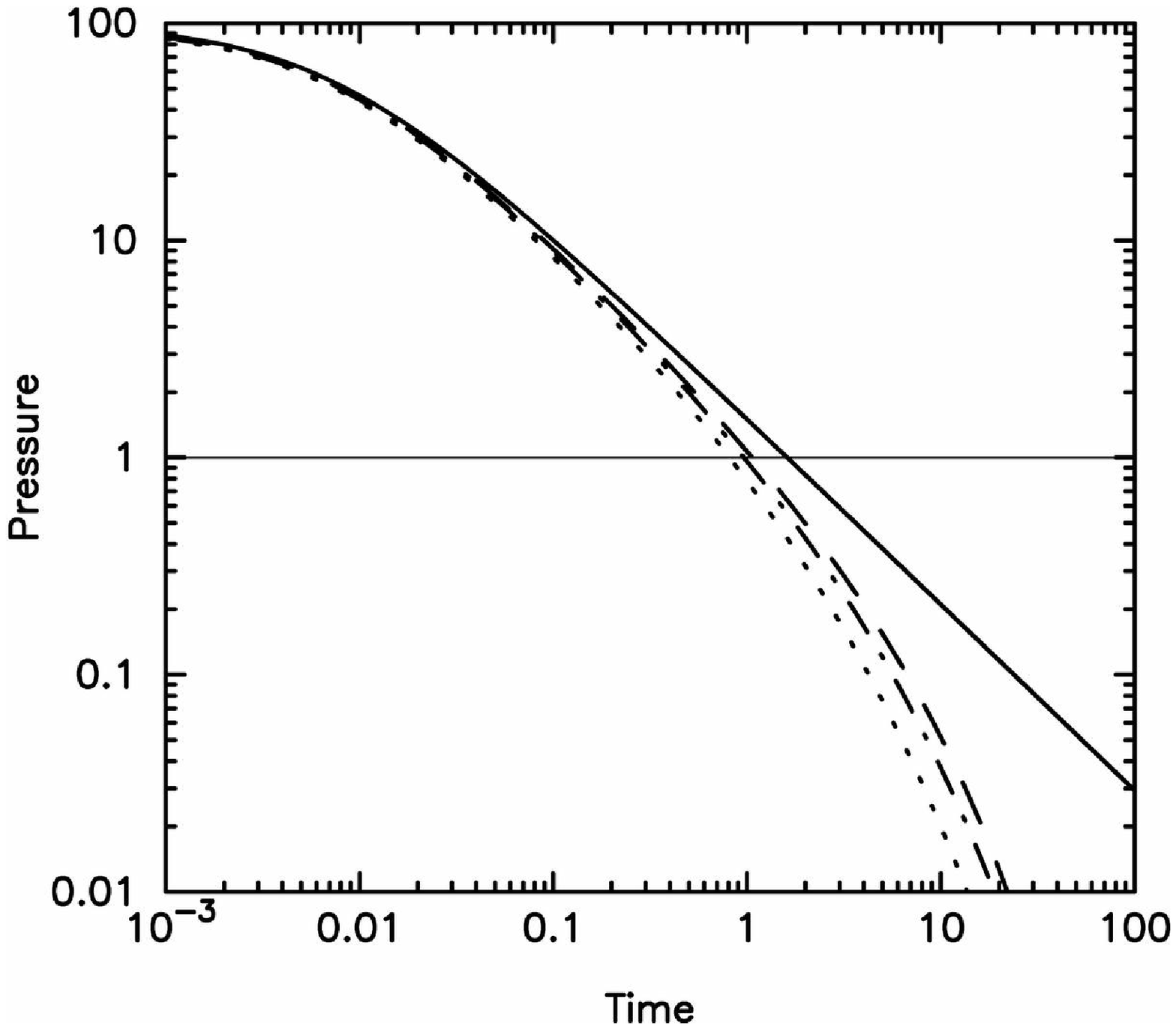}
\caption{Internal to external pressure ratio against time (in units of
  the sound crossing time). \textit{Top}: $\mu = 1$
  direction. \textit{Bottom}: $\mu = -1$ direction. Linetypes same as
  in Fig.~\ref{fig:radupdown}. The horizontal line marks equal
  pressures.}
\label{fig:presupdown}
\end{figure}

In Figure~\ref{fig:presupdown} we show the ratio of the pressure in
the ionized gas to that in the ambient medium, assuming that the
latter is isothermal with a temperature of 100~K. From these figures
we see that in the $\mu = -1$ direction the pressures equalize in
about 1 sound crossing time for both $\alpha = 2$ and $\alpha = 3$. In
the $\mu = 1$ direction, the $\alpha = 3$ case shows that the
pressures do not equalize before the ionization front begins to
accelerate. In the $\alpha = 2$ case, the pressures do equalize at
around 7 sound crossing times for the adopted parameters (i.e.,
initial Str\"omgren radius, $R_0$). We remark that once the pressures
become similar, the analysis of this section does not really hold,
since the external pressure should be taken into account. However, we
also comment that the fact that the pressures do become equal before
the shock begins to accelerate in the $\alpha = 2$ case means that, in
practice, the \ion{H}{2} region will remain confined for this choice
of initial Str\"omgren radius.

\begin{figure}[!t]
\includegraphics[width=0.45\textwidth]{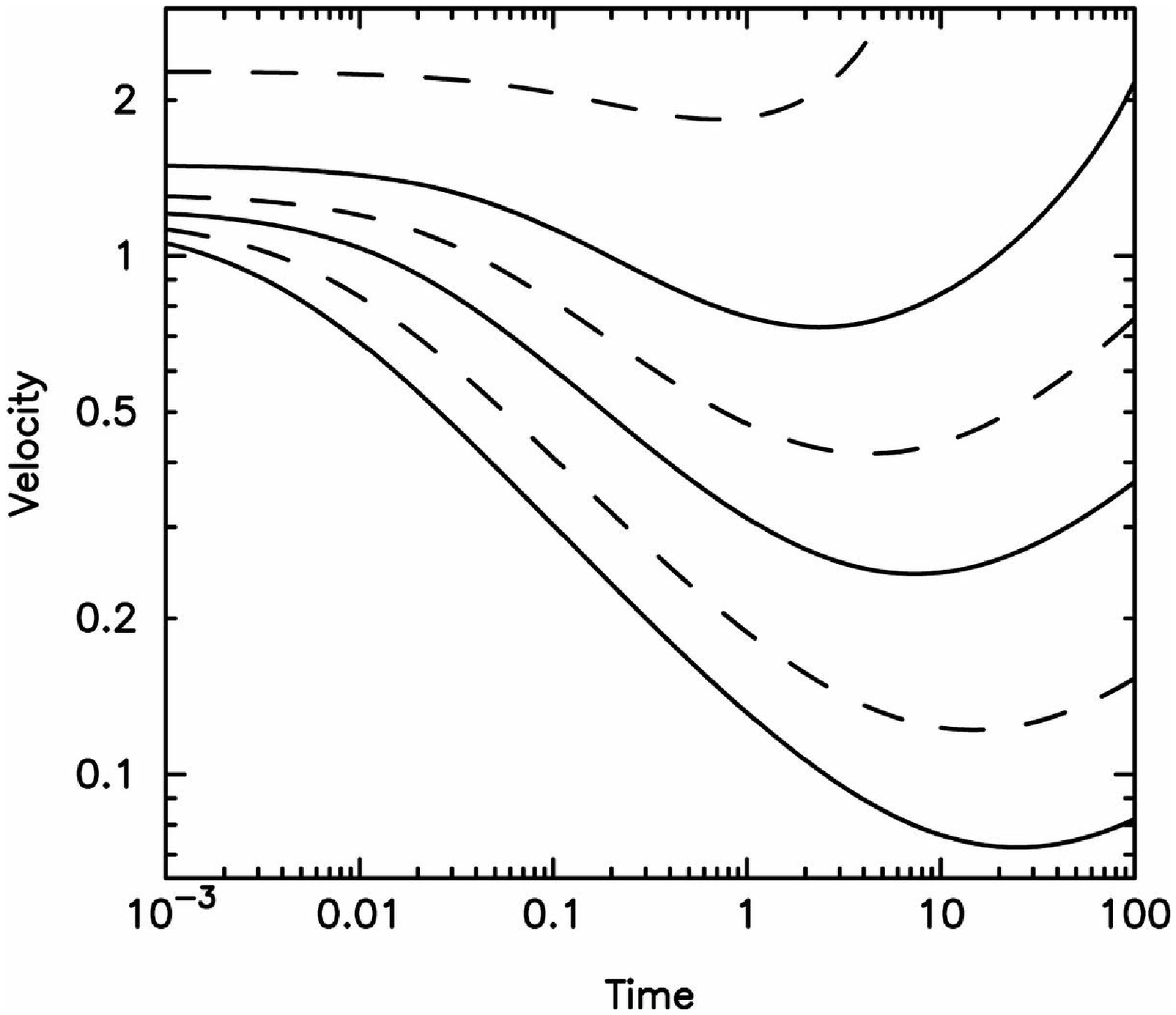}\\
\includegraphics[width=0.45\textwidth]{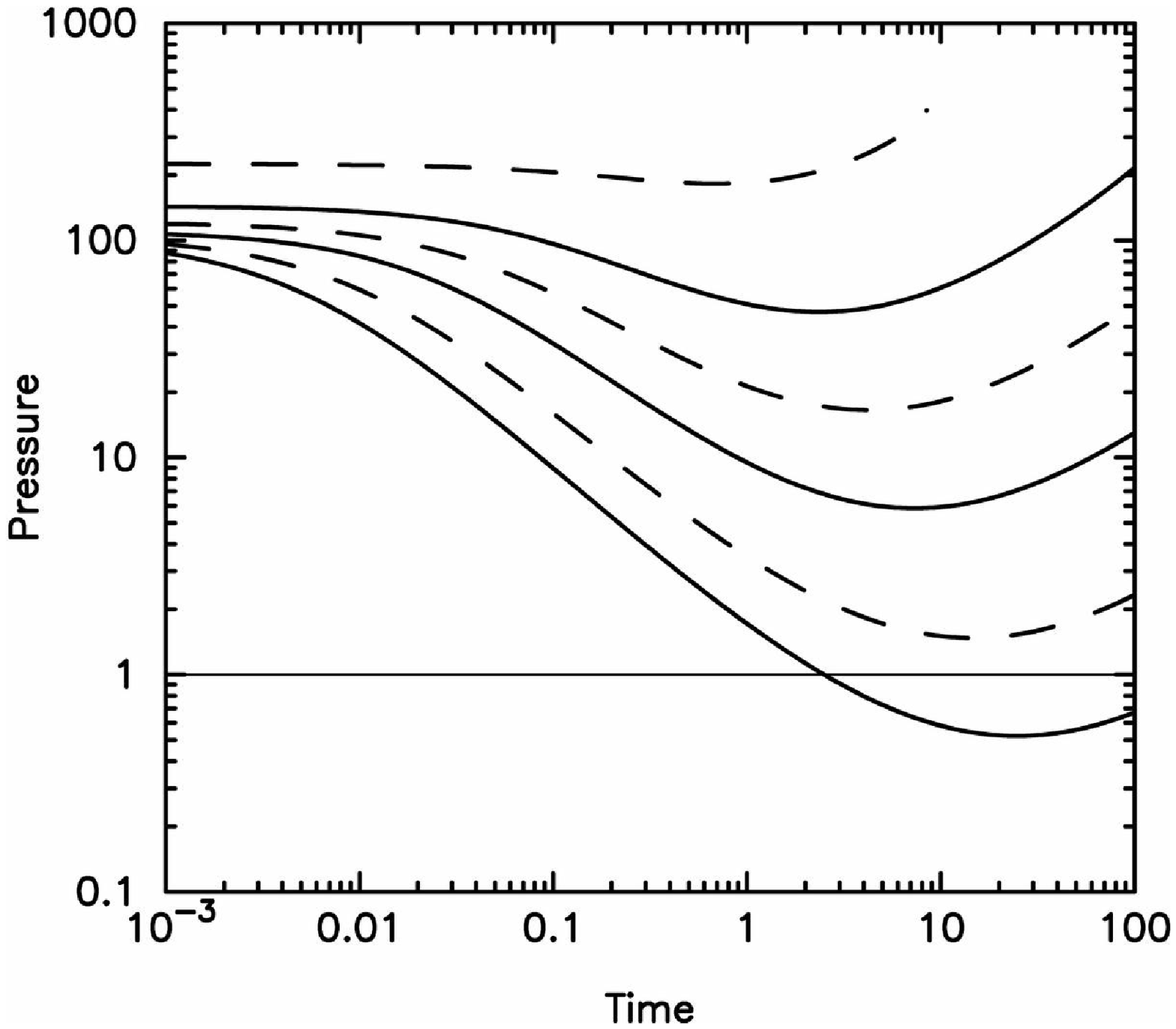}
\caption{\textit{Top}: Velocity of the ionization front (in units of
  the ionized gas sound speed) against time (in units of the sound
  crossing time) for different values of the initial Str\"omgren
  radius. From lowest line to uppermost line: $R_0/r_c = 0.01$, 0.02,
  0.05, 0.1, 0.2, 0.5. \textit{Bottom}: Ratio of internal to external
  pressures for the same initial Str\"omgren radius values. The
  horizontal line denotes equal pressures.}
\label{fig:multidown}
\end{figure}
In Figure~\ref{fig:multidown} we show how the breakout time of the
ionization front in the $\mu = 1$ direction varies
with initial Str\"omgren radius, for values of $R_0/r_c$ between 0.01
and 1.0 for the case $\alpha = 2$. As $R_0/r_c \rightarrow 1$ the
position of the velocity minimum moves to earlier times and higher
velocities. Also in this figure, we show how the internal to external
pressure ratio varies with initial Str\"omgren radius. Only for $R_0/r_c
\lesssim 0.02$ is there a regime for which pressure balance could
possibly halt the expansion and subsequent breakout of the \ion{H}{2}
region. 

\section{Numerical Simulation}
\label{sec:numerical}
\begin{figure}[!t]
\includegraphics[width=0.45\textwidth]{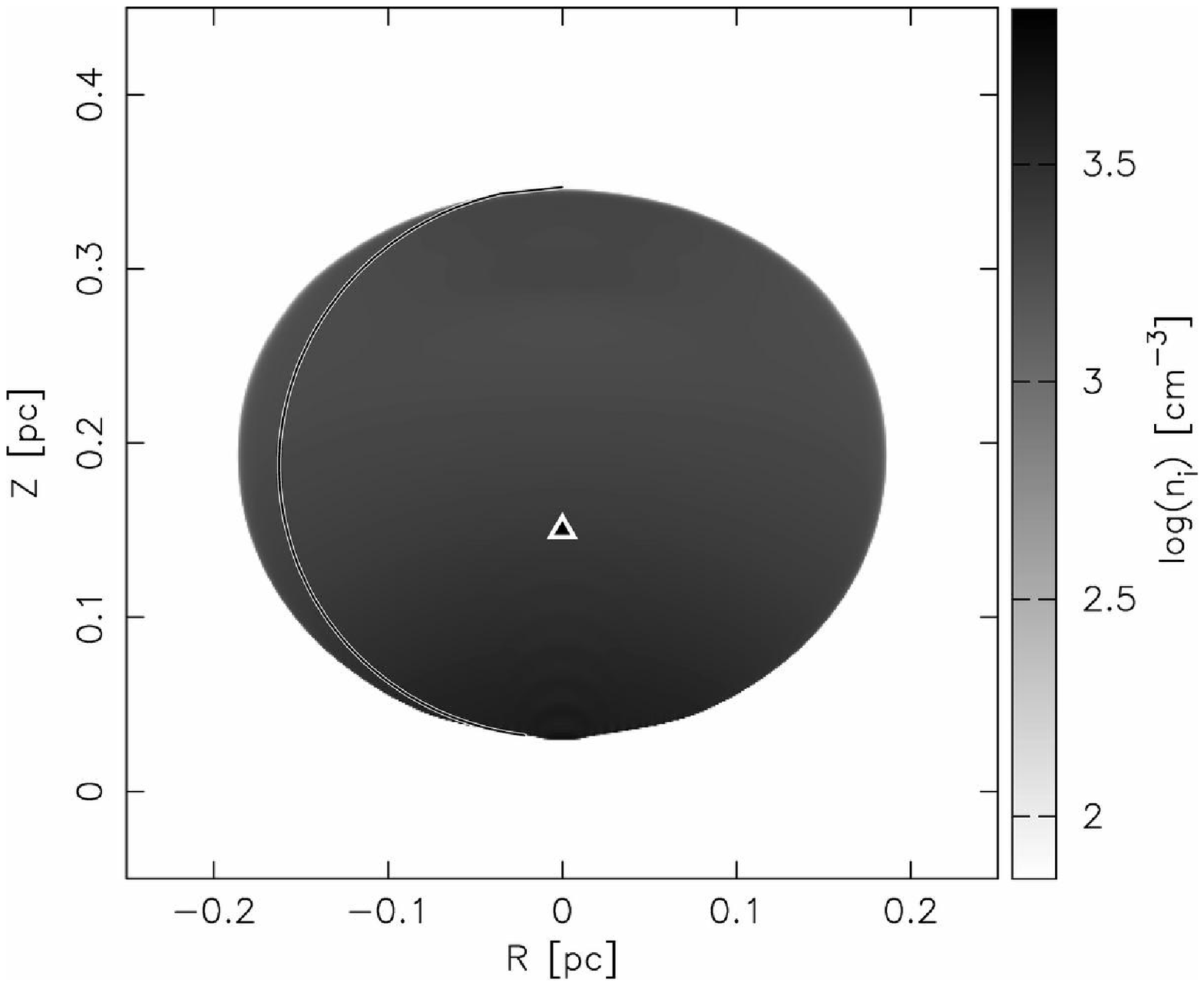}\\
\includegraphics[width=0.45\textwidth]{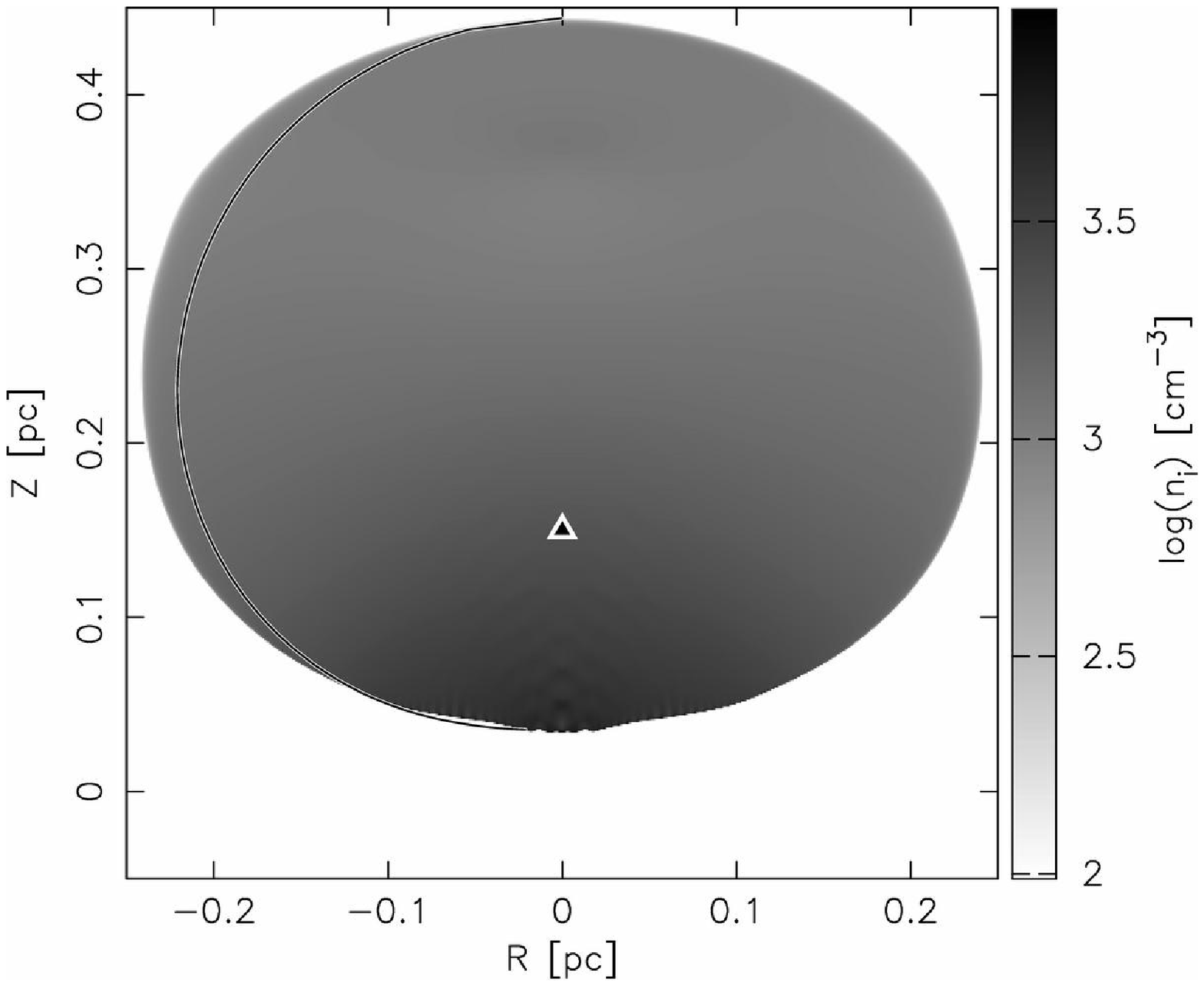}
\caption{Logarithm of ionized gas density. \textit{Top}: $\alpha =
  2$ numerical solution after 60,000~yrs. \textit{Bottom}: $\alpha = 3$
  numerical solution after 50,000~yrs. The black triangle marks the
  position of the star and the position $(0,0)$ denotes the center of
  the density distribution in each case. The heavy black curve
  represents the shape of the ionization front obtained from the
  approximation of Equation~\ref{eq:radius}. }
\label{fig:numden23}
\end{figure}

In order to test our assumption that the density within the ionized
gas is approximately constant during the expansion stage, we perform
numerical simulations of the expansion of an off-center \ion{H}{2}
region in a power-law density distribution. The density law for this
simulation is taken from \citet{2005arXiv.0508467.F}, namely
\begin{equation}
  n(r) =  \left\{
\begin{array}{ll}
 n_0  \exp\left[-\frac{\alpha}{2}\left(r/r_c\right)^2\right] & \mbox{for
  $r \leq r_c$} , \\
n_0 e^{-\alpha/2} \left( r/r_c \right)^{-\alpha} & \mbox{for $r \geq
  r_c$} , 
\end{array}
\right.
\label{eq:numsimden}
\end{equation}
where $n_0$ is the central density and $r_c$ is the stellar offset
distance from the center of the density distribution. At $r_c$ the
density has the value $n_c \equiv n(r_c) = n_0/e^{\alpha/2}$.

This density distribution is flatter for radii $r < r_c$ than the
simple power-law density distribution discussed in \S\S~\ref{sec:prob} and
\ref{sec:expand} above. It is obtained by assuming that the
acceleration due to gravity of the hydrostatic equilbrium solution in
the halo, $r > r_c$, matches that of the core (defined as $r < r_c$)
at $r = r_c$, with the added requirement that the acceleration due to
gravity must be zero at the center of symmetry. This particular
density law assumes that the acceleration due to gravity is linear
with radius for $r < r_c$. We adopt this density distribution rather
than a simple power law because the density does not go to infinity at
the center of symmetry and the ``core-halo'' structure is more
representative of real molecular clouds. 

Although we do not expect the ionization front expansion of our
numerical simulations to agree with the analytic expansion found in
\S~\ref{sec:expand} in the region between the stellar position and the
center of the density distribution (i.e., $r < r_c$), because the
density laws are very different here, we do expect to be able to make
a direct comparison in the halo region, $r > r_c$, where the density
laws are the same. Since we are primarily interested in whether the
\ion{H}{2} region remains bounded in the $\mu = 1$ direction during
the expansion stage, then it is sufficient that the density
distributions are the same in this direction for radii $r > r_c$.

The radiation-hydrodynamics code employed in these simulations has
been described previously in \citet{2006ApJS..165..283A}. The model
parameters specific to this work are a central density $n_0 = 3.5
\times 10^5$~cm$^{-3}$, and distance from density center to the star
of $r_c = 0.15$~pc. We assume isothermal, hydrostatic equilibrium in
the initially neutral 100~K gas, and include an external gravity
field (see \citealp{2005arXiv.0508467.F} for details). The
gravitational acceleration does not have any noticeable dynamical
effect on the expansion of the \ion{H}{2} region.  The ionizing star
produces $10^{48}$~photons~s$^{-1}$. The grid size for these
calculations is $250 \times 600$ cells, representing a spatial size of
$0.25 \times 0.6$~pc. For these parameters, $y_{\rm sc} = R_0/r_c =
0.08235$ for the $\alpha = 2$ case and $y_{\rm sc} = R_0/r_c =
0.11492$ for the $\alpha = 3$ case.
\begin{figure}[!t]
\includegraphics[width=0.45\textwidth]{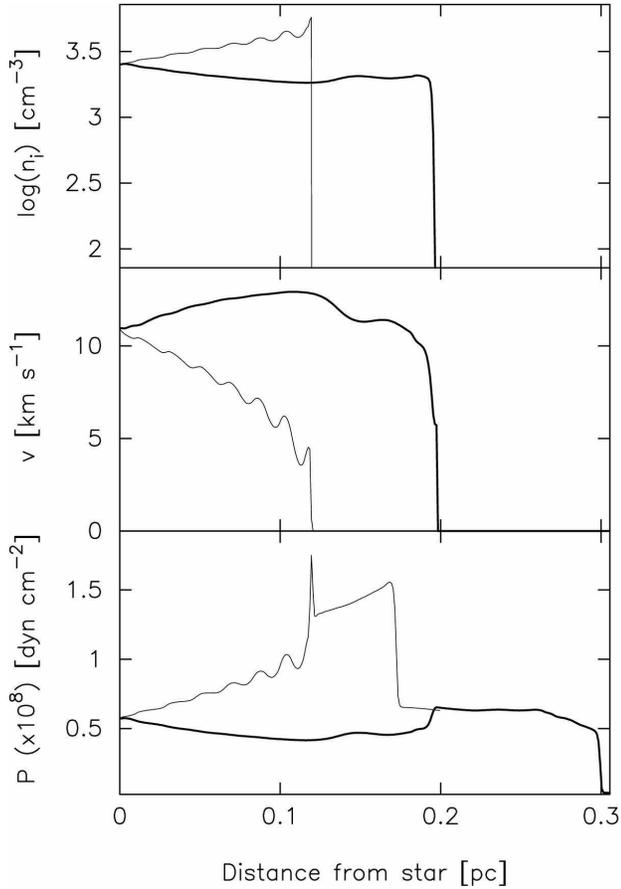}
\caption{Cuts along the $\mu = \pm 1$ axis of the $\alpha = 2$
  numerical simulation shown in Fig.~\ref{fig:numden23}.
  \textit{Top}: density. \textit{Middle}: velocity.
  \textit{Bottom}: pressure. The thick solid line represents the $\mu
  = 1$ direction and the thin solid line represents the $\mu = -1$
  direction. Units are as indicated on the axes.}
\label{fig:numcut2}
\end{figure}
\begin{figure}[!t]
\includegraphics[width=0.45\textwidth]{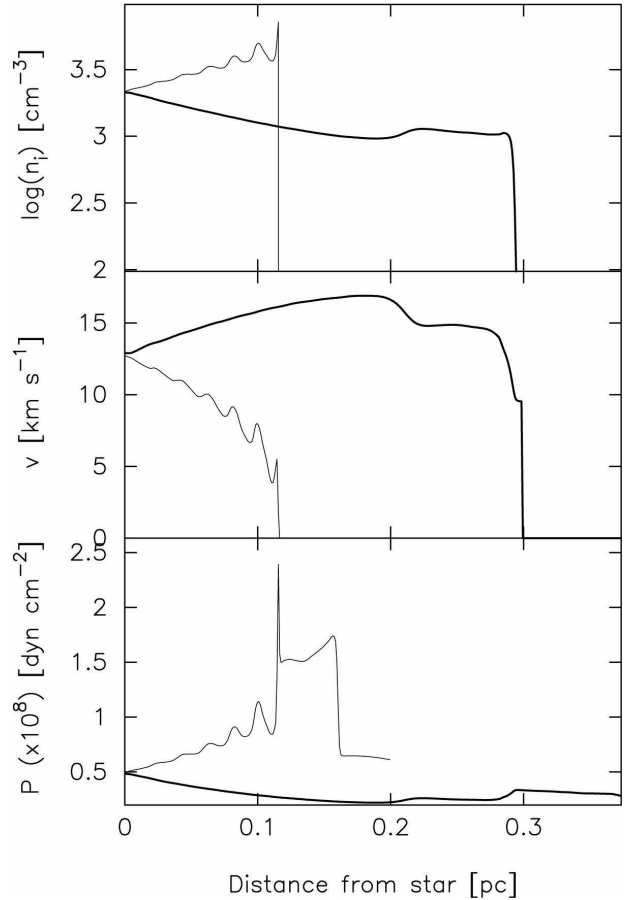}
\caption{Cuts along the $\mu = \pm 1$ axis of the $\alpha = 3$
  numerical simulation shown in Fig.~\ref{fig:numden23}.
  \textit{Top}: density. \textit{Middle}: velocity.
  \textit{Bottom}: pressure. The thick solid line represents the $\mu
  = 1$ direction and the thin solid line represents the $\mu = -1$
  direction. Units are as indicated on the axes.}
\label{fig:numcut3}
\end{figure}

In Figures~\ref{fig:numden23}, \ref{fig:numcut2} and \ref{fig:numcut3}
we show the results of calculations of off-center \ion{H}{2} regions
in $\alpha = 2$ and $\alpha = 3$ power-law density
distributions. Figure~\ref{fig:numden23} shows the ionized gas
density, i.e. the shape of the \ion{H}{2} regions, for both power laws
after 60,000 and 50,000~yrs respectively. The density within the
ionized regions is approximately constant. We note that for both
models, the ionized volumes are approximately spherical, although the
star is off-center. This was also noted by \citet{2006arXiv.0605501.M}
in their three-dimensional numerical simulations. Overplotted on
Figure~\ref{fig:numden23} is the shape the ionization front would have
according to the approximation of Equation~\ref{eq:radius}. We can see
that the approximation slightly underestimates the volume of the
ionized gas with respect to the numerical simulations.

In Figures~\ref{fig:numcut2} and \ref{fig:numcut3} we show cuts along
the symmetry axis, i.e. the $\mu = \pm 1$ directions, for both
cases. From these figures we see that the assumption of constant
density is reasonable for the $\mu = 1$ (``downhill'') direction but
not so good in the $\mu = -1$ direction, since the density increases
with distance from the star. The pressure follows the same pattern as
the density, as is to be expected in for an isothermal, $T = 10^4$~K
\ion{H}{2} region. Outside the ionized region, in both directions, the
pressure is high because the neutral shock sent out ahead of the
ionization front pressurizes the gas. The velocities in the ionized
gas are subsonic in the $\mu = -1$ direction and become supersonic in
the $\mu = 1$ direction.

The ripples in the $\mu = -1$ subsonic gas are due to the ionization
front not being well resolved in this case, since the high densities
lead to a very thin transition zone. The spike in the pressure is due
to the heating and cooling being out of equilibrium at the unresolved
ionization front. These numerical problems are not dynamically
important for these simulations, since we are interested in the
expansion in the $\mu = 1$ direction, where the gas is supersonic and
hence not affected by fluctuations in the thermal pressure. This
problem was noted by \citet{2005ApJ...627..813H}. Their solution to
the problem was to reduce the effective ionization cross section,
$\sigma_0$, by a factor of 60. In this paper, we reduce $\sigma_0$ by
a factor of 10. Reducing the ionization cross section makes the
ionization front broader and so allows it to be resolved by the
numerical scheme. Since the width of the ionization front is of the
order of a few photon mean free paths, where $l_p \propto (n
\sigma_0)^{-1}$, then the ionization front is narrower in high density gas
than in low density gas, and hence more of a problem to resolve
numerically in the former.  In the $\mu = 1$
direction, the much lower densities mean that the ionization front is
easily resolved by the numerical scheme.
\begin{figure}[!t]
\includegraphics[width=0.45\textwidth]{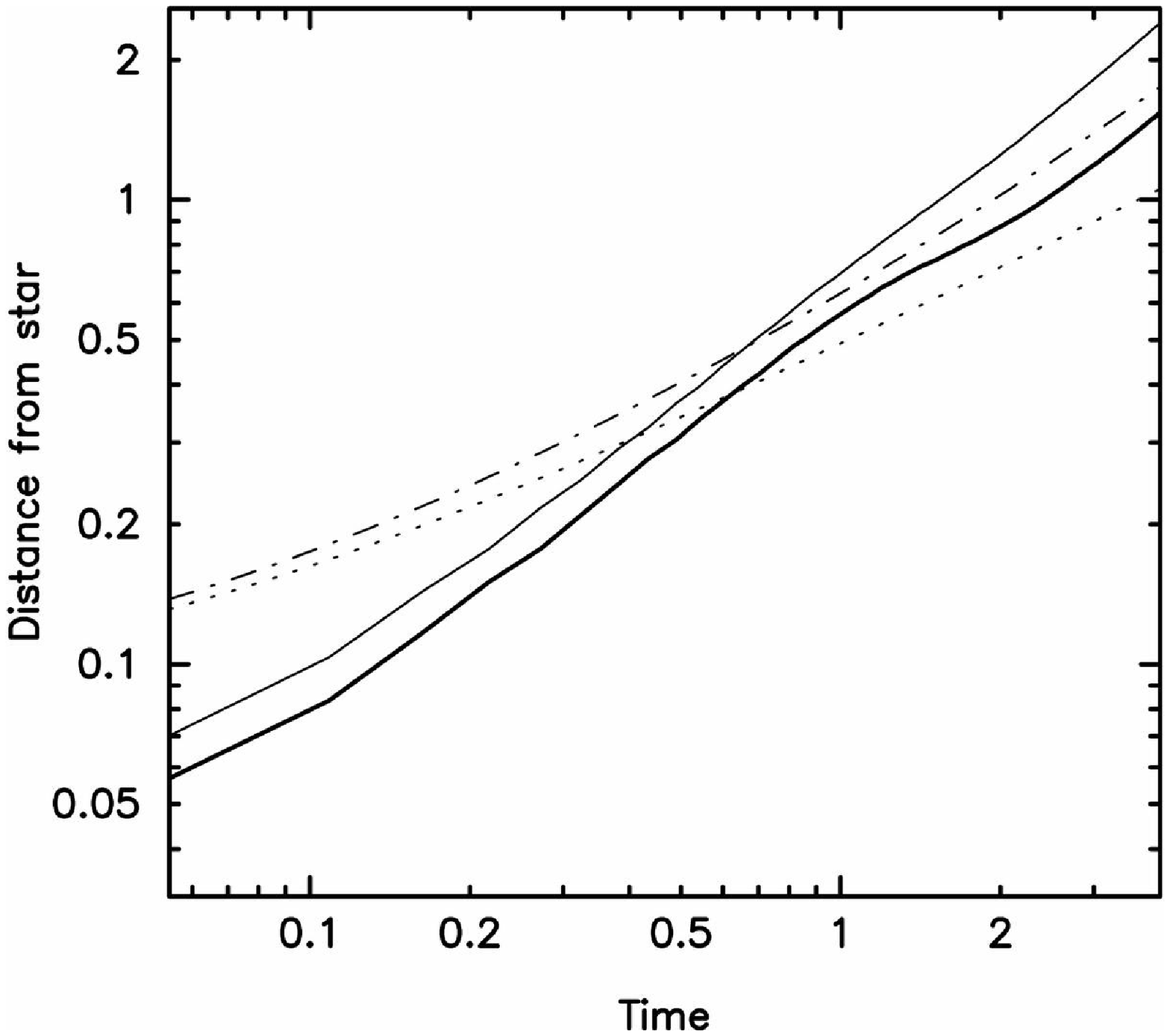}\\
\includegraphics[width=0.45\textwidth]{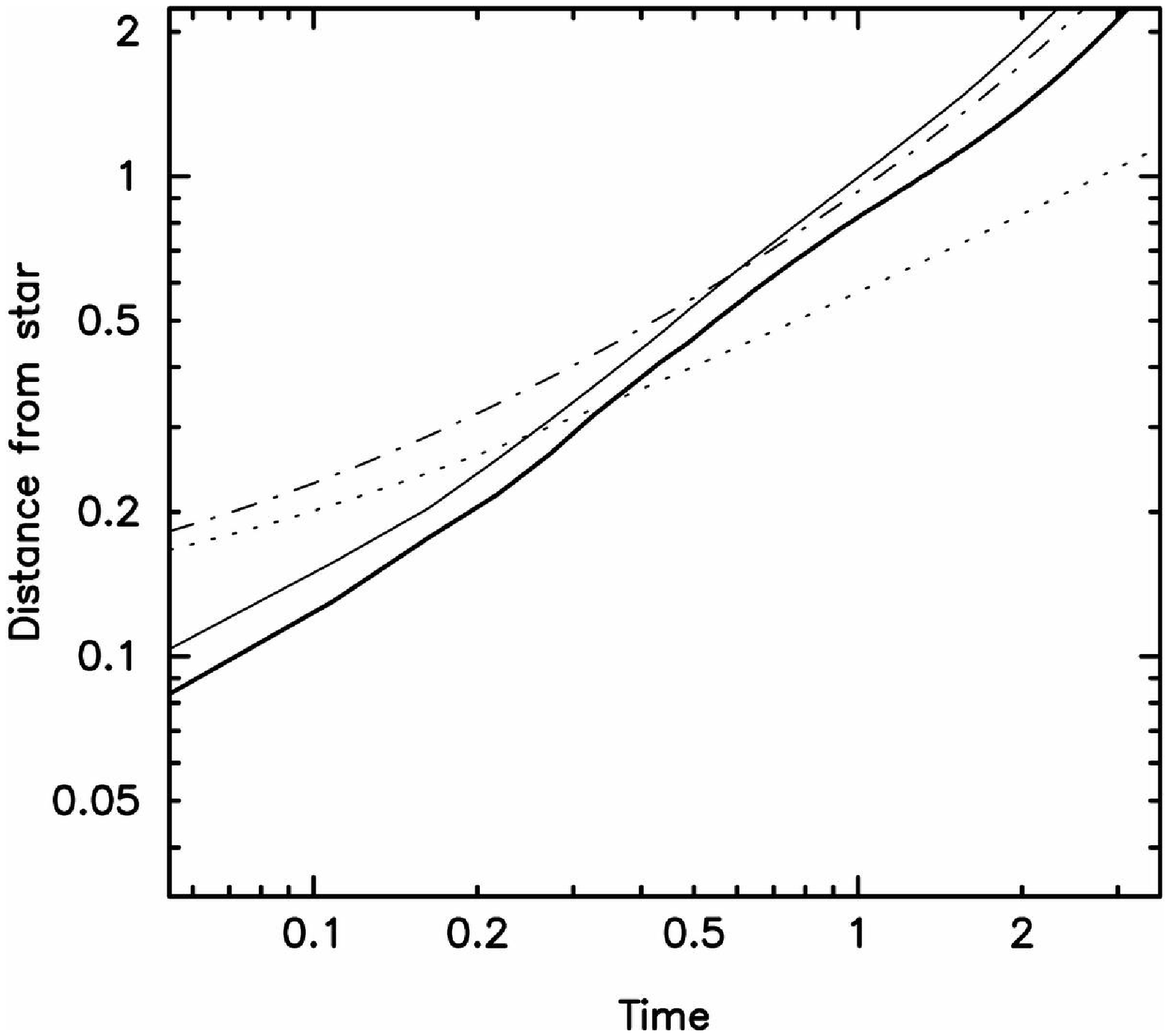}
\caption{Radius (in units of stellar offset radius) against time (in
  units of sound crossing time) in the $\mu = 1$ direction, showing ionization
  front radius (\textit{thick solid line}), shock front radius
  (\textit{thin solid line}), analytical solution for the same
  parameters (\textit{dot-dashed line}) and uniform density analytical
  solution (\textit{dashed line}).
  \textit{Top}: $\alpha = 2$ numerical simulation and corresponding
  analytical solution. \textit{Bottom}: $\alpha = 3$ numerical
  simulation and corresponding analytical solution.}
\label{fig:numanal23}
\end{figure}

In Figure~\ref{fig:numanal23} a comparison of the numerical simulation
and the analytic solution for the same parameters in the $\mu = 1$
direction is plotted. Here, the distance scale is in units of the
Str\"omgren radius in a medium of uniform density $n_c e^{-\alpha/2}$,
and time is given in units of the sound crossing time of the
Str\"omgren sphere. Since the density distribution in the $\mu = -1$
direction of the numerical simulation is different to the analytical
model discussed in Section~\ref{sec:expand}, we do not show a
comparison graph for this case, as it is meaningless. Instead, we
focus on the $\mu = 1$ direction, for which the numerical simulation
and analytical model density distributions are the same. We plot both
the shock wave and the ionization front trajectories, together with
the analytical result. For times less than about half a sound crossing
time, there is not a good agreement between analytical and numerical
results. This can be partly attributed to resolution effects, since
the initial Str\"omgren radius in this high density medium is very
small, and the details of the hydrodynamics are not particularly well
resolved at this scale. For times greater than half a sound crossing
time, the agreement between numerical and analytical solutions is very
close for the remainder of the calculation.
The analytical solution falls between the shock front and ionization
front paths for the whole of this time. 

We do not follow the numerical simulation beyond around 4 sound
crossing times because part of the ionized gas volume leaves the
computational grid at this point.

\section{Summary}
\label{sec:summary}
In this paper we have developed an analytical treatment for the
formation and expansion of an \ion{H}{2} region off center in a steep
power-law density distribution. We find that during the initial
formation stage, the \ion{H}{2} region will remain bounded as long as
the ratio of the Str\"omgren radius in the equivalent uniform medium
at the stellar position to the stellar offset position is less than a
value that depends only on the power-law exponent (see
Eq.~\ref{eq:root}). In the expansion phase, the ionization front will
break out (turn supersonic with respect to the ionized gas) unless
pressure balance between the internal photoionized gas and the
external medium is achieved before this point. For Str\"omgren radius
to stellar offset radius ratios of more than 0.02, pressure balance is
not possible for an isothermal 100~K medium in hydrostatic equilibrium
and so the \ion{H}{2} region will become unbounded, eventually, in
these cases.

We performed numerical simulations to test the assumption of constant
density within the ionized gas during the expansion phase adopted in
\S~\ref{sec:expand}. We found that this assumption is valid in the
``downhill'' ($\theta = 0$) direction from the star, but is not so
good in the opposite (``uphill'', $\theta = \pi$) direction. In the
$\theta = 0$ direction, the numerical solution and analytical solution
follow the same expansion law once the elapsed time is greater than
about half a sound crossing time. We also assessed the validity of the
simple analytical approximation to the ionization front
shape. Although the volume of ionized gas is slightly underestimated
in the analytical approximation, the effect is not significant, as
evinced by the agreement between the numerical and analytic expansion
laws. This gives us confidence that our simple description of the
\ion{H}{2} region expansion can provide insight into this complicated
problem.

\acknowledgments We thank Will Henney for a critical reading of the
manuscript and Pepe Franco and Stan Kurtz for valuable comments. SJA
would like to acknowledge support from DGAPA-UNAM through PAPIIT grant
IN112006. This work has made use of NASA's Astrophysics Abstract Data
Service and the astro-ph archive.

\end{document}